\documentclass[a4paper]{amsart}

\usepackage{lineno}
\usepackage{amssymb}
\usepackage[final]{showkeys}
\usepackage{microtype}
\usepackage{color}
\usepackage{graphicx}
\usepackage{dcolumn}
\usepackage[hmargin=4.5cm,vmargin={4cm,4cm}]{geometry}

\numberwithin{equation}{section}

\allowdisplaybreaks

\begin{document}

    \title[Fractional conservation of mass and Gas Flow Problems]{
    Application of the fractional conservation of mass to Gas Flow diffusivity equation in heterogeneous porous media}

	\author{Arrigo Caserta$^1$}
	        \address{Istituto Nazionale di Geofisica e Vulcanologia, Roma}	

	\author{Roberto Garra$^2$}
        \address{Dipartimento di Scienze Statistiche, ``Sapienza'' Universit\`a di Roma.}

	\author{Ettore Salusti$^3$}
        \address{INFN, Sezione Roma 1.}

    \date{\today}

    \begin{abstract}
     In this paper we reconsider the classical nonlinear diffusivity 
     equation of real gas in an heterogenous porous medium in light of the recent studies about the generalized fractional equation of conservation of mass.
     We first recall the physical meaning of the fractional conservation of mass recently studied by Wheatcraft and Meerschaert \cite{mark} and then consider the implications in the classical model of diffusion of a real gas in a porous medium. Then we show that the obtained equation can be simply linearized into a classical space-fractional diffusion equation, widely studied in the literature.
     We also consider the case of a power-law pressure-dependence of the permeability coefficient. In this case we provide some useful exact analytical results. In particular, we are able to find a Barenblatt-type solution for a space-fractional Boussinesq equation, arising in this context.
    \end{abstract}

    \maketitle

    \section{Introduction}
    
    Mathematical modelling of gas flow propagating in porous media involves a remarkable number of difficulties, 
    due to the possible variability of permeability and the interaction of the fluid with the medium. Moreover in some cases, the heterogeneous shape of the porous medium should be taken into account. In recent papers some discussions about the limit of validity of classical conservation of mass and Darcy law have been proposed by several authors (see for example \cite{cap,ochoa,mark} and references therein). In particular, in various 
    linear models of the physics of solid earth, fractional derivative operators have been considered as useful mathematical tools to consider memory effects and trapping (by means of fractional derivatives in time) and nonlocal behavior and jumps with long
    tails (fractional derivatives in space). In particular fractional advection-dispersion equations have been widely studied
    for modeling transport processes at the Earth surface, we refer to the relevant paper by Schumer et al. \cite{schumer} about this topic.
    
     From the macroscopic point of view, time-fractional models in the physics of porous media can be heuristically based on a modified Darcy law with memory that takes into account the change in time of porosity of the medium in the interaction with the fluid (see e.g. \cite{cap} and \cite{moroni}). Space-fractional models can be derived by using a generalized fractional conservation of mass, according to the approach developed by Wheatcraft and Meerschaert in \cite{mark}, where fractionality results from the heterogeneity of the control volume. The most of the studies on this topic regards
     linear models, where nonlinear terms are completely neglected, even if they play a relevant role in many realistic problems. On the other hand 
     linear space-time fractional models are completely motivated from the probabilistic and microscopic point of view (see e.g. \cite{schumer}), while 
     physical discussions about nonlinear models involving fractional derivatives are almost missing. We observe that in the recent mathematical 
     literature many papers have been devoted to the study of different versions of the space-fractional porous medium equation (for example \cite{vaz} and many others) and relevant results have been raised out, but the physical counterpart of these studies is in our view still inadequate.
     A derivation of nonlinear fractional models from modified physically based constitutive laws can help to better understand their utility 
     for applications. Moreover the investigation about exact solutions in this framework can play a key-role to understand 
     some relevant features, such as finite velocity of propagation or blow-up in finite time and so on.
     
    As far as we know, the application of the generalized law of conservation of mass to realistic non-linear models  of fluid flow 
    in heterogenous porous media  have not been still considered in the literature. The main purpose of this paper is to reconsider the classical diffusivity equation of real gas in an heterogeneous porous medium in light of the recent discussions about the fractional conservation of mass.
    We will firstly derive the space-fractional diffusivity equation, starting from the fractional conservation of mass introduced by Wheatcraft and Meerschaert in \cite{mark} and then show that, under suitable assumption, this equation is simply linearizable to a linear space-fractional diffusion equation, widely studied in the literature.
    
    Then we will consider the fractional diffusivity equation in the case of pressure-dependent permeability. In particular we analyze the high pressure regimes when the permeability growth with pressure according to a power law relation. In this way
    an higher order nonlinear term appears and we obtain by change of variable a fractional porous-medium type equation.
    In this case we provide several new explicit results by using a generalized separating variable method and 
    the invariant subspace method. In this context, we are able to find a Barenblatt's type solution for the space-fractional Boussinesq-like equation.
    While in the linear fractional diffusion equation, the real order of the space-fractional derivative has a clear physical meaning in the
    framework of CTRW models (and therefore space-fractional diffusion models have been experimentally validated in various contexts \cite{metz}), in the nonlinear case 
    an experimental validation is still missing. We discuss, by means of the Barenblatt's type solution, the role played by the real order of the space-fractional derivative
    on the velocity propagation of the pressure front. Thus we suggest the way to experimentally evaluate the order of the fractional
    derivative starting from the measurement of the front propagation velocity.\\
    This result for the space-fractional porous medium type equation can have relevant applications for example in the context of 
    the studies about fluid-induced microseismicity \cite{shapiro, sh1}. Moreover, in Appendix C, we show that the real gas space-fractional diffusivity    
    equation can be reduced to the porous medium-type equation in the more general case of pressure dependent coefficients, under suitable 
    mathematical assumptions. This observation stresses the utility of the obtained results in a really general unitary theory, 
    where both pressure-dependence of the physical coefficients and nonlocality due to heterogenity are taken into account.
     
    The main aim of this paper is 
    finally to provide a critical revisitation of classical diffusivity models, in light of recent studies appeared in the literature
    about the relation between fractional derivative and mass flux in heterogeneous media. A second concrete 
    outcome is given by the discussion of some new explicit analytical results that are related to interesting 
    features of the generalized diffusivity equation. 
    This study represents in our view a first step in the analysis of non-linear non-local models in physics of solid earth. 
    
    The paper is organized as follows: in Section 2 a short survey about the derivation of the space-fractional conservation of mass is provided and in Section 3 we discuss its application to the real gas diffusivity model. In Section 4 we study the case of space-fractional diffusivity equation with pressure-dependent permeability and we show some exact analytical results, including a Barenblatt's type solution that can help to experimentally evaluate the real order of the fractional derivative. Three Appendices are also presented, about fractional derivatives, the invariant subspace method and the mathematical treatment of a more general space-fractional diffusivity equation.

    \section{The fractional conservation of mass: a short survey}
    
    In the recent paper \cite{mark}, the authors have discussed the limit of validity of the classical fluid mass conservation 
    equation, starting from its mathematical derivation. We here briefly recall their reasoning starting from the derivation 
    of this classical equation in hydrology.
    
    Let us consider a control volume with lenght of the sides $\Delta x_1$, $\Delta x_2$ and $\Delta x_3$. The inflow
    component passing through the $-x_1$ face is given by
    \begin{equation}
    F(x_1)= \Delta x_2 \Delta x_3 \rho q_1,
    \end{equation}
    where $q_1$ is the $x_1$-component of the flux and $\rho$ the fluid density.
    The mass flux outflow through $+x_1$ is given by
    \begin{equation}
    F(x_1+\Delta x_1)= \Delta x_2 \Delta x_3 \rho q_1+\Delta x_2 \Delta x_3 \frac{\partial \rho q_1}{\partial x_1}\Delta x_1,
    \end{equation}
    under the assumption that a first-order Taylor approximation for the mass flux expanded about $x_1$ can be taken.
    From the physical point of view, this approximation corresponds to the assumption that changes in mass flux within 
    the control volume are linear. In this case the net mass flux in the $x_1$ direction is given by
    \begin{equation}
    F(x_1)-F(x_1+\Delta x_1) = -\Delta x_2 \Delta x_3 \frac{\partial \rho q_1}{\partial x_1}\Delta x_1 = -\Delta V \frac{\partial \rho q_1}{\partial x_1}.
    \end{equation}
    With the same reasoning, by calculating the flux on the other directions, we obtain the divergence of the net flux of mass
    and finally the standard mass conservation for a fluid in a porous medium, that is
    \begin{equation}
    \frac{1}{\Delta V}\frac{\partial}{\partial t}\rho \phi \Delta V= \nabla \cdot
        \rho \mathbf{q},
    \end{equation} 
    where $\phi$ is the porosity and $\mathbf{q}= (q_1(\mathbf{x},t),q_2(\mathbf{x},t),q_3(\mathbf{x},t))$.
     
    The limit of this classical form of the conservation of mass, as pointed out by Wheatcraft and Meerschaert in \cite{mark}, is given by
    the fact that it requests two main physical assumptions: 
    \begin{itemize}
    \item the flow through the control volume should be linear, 
    \item the control volume must be large compared to the heterogeneity scale.
    \end{itemize} 
    Under these assumptions, the mathematical treatment of the problem of mass flux through a control volume can be based on 
    a first order Taylor approximation, as just seen. Indeed, in this case 
    the control volume vanishes at the limit and a first order expansion can be applied.\\
    These assumptions should be relaxed in many physical cases. In particular, for heterogeneous porous media,
    the control volume is scale-dependent and will be much larger than the measurement scale. In this case the first-order approximation
    in the mathematical derivation of the continuity equation is not still valid and a different approach is requested.
    
    The approach suggested by Wheatcraft and Meerschaert in \cite{mark}, is based on the application of a fractional Taylor series 
    expansion for the mass flux through an heterogeneous control volume.
    They used in the mathematical derivation of the fractional conservation of mass the fractional Taylor expansion introduced in the
    literature by Odibat and Shawagfeh \cite{Odibat}. According to their definition 
    the fractional Taylor expansion of the mass flow $F$ in the point $y= x+\Delta x$ is given by
    \begin{align}
    \nonumber & F(y)= F(x)+D_x^\gamma F(x^+)\frac{(y-x)^\gamma}{\Gamma(\gamma+1)}+D_x^\gamma D_x^\gamma F(x^+)\frac{(y-x)^{2\gamma}}{\Gamma(2\gamma+1)}+\dots ,
    \end{align}
    where 
    \begin{equation}
    D_x^\gamma F(y)= \frac{1}{\Gamma(1-\gamma)}\int_x^y \frac{F'(y-u)}{(u-x)^\gamma}du,
    \end{equation}
    is the Caputo fractional derivative of order $\gamma \in (0,1)$ (see \cite{kil} and Appendix A). By using this approach a non-linear power law flux through an highly heterogeneous medium can be studied. 
   In this paper we consider a porous solid as an highly heterogeneous medium when pore size and shape are strongly variable. Even if this is an heuristic definition, on a laboratory scale it can be seen as a characteristic of the solid due to variation in cobble size and particule distribution of sand grains.
    
    We refer to the enlighting paper of Wheatcraft and Meerschaert for the full derivation of the fractional conservation of mass, based on these heuristic arguments and mathematical tools. The resulting governing equation for fractional conservation of mass
    is given by
    \begin{equation}\label{fmas}
    \frac{\partial}{\partial t}\rho \phi = \nabla^\gamma \cdot
    \mathbf{q},
    \end{equation}
	where the fractional divergence of the flux is given by
	\begin{equation}
	\nabla^\gamma \cdot
	    \mathbf{q} = \frac{\partial^\gamma q_1}{\partial x_1^\gamma}+
	    \frac{\partial^\gamma q_2}{\partial x_2^\gamma}+
	    \frac{\partial^\gamma q_3}{\partial x_3^\gamma}.
	\end{equation}    
     Therefore in this physical derivation of the fractional conservation of mass the meaning of the real order of the derivative $\gamma$ is related to the geometry of the medium that implies a physical non-locality as an effect of the heterogeneity of the porous matrix. 
   From an experimental point of view, the real parameter $\gamma$ can be measured as a deviation factor with respect to the classical flux law.
    
    We notice that a more general treatment of the fractional generalization of the conservation of mass has been recently discussed in \cite{olsen}. In that paper a two--sided fractional conservation of mass equation was derived by using left and right fractional Mean Value Theorems and the one-sided approach 
    discussed in \cite{mark} was recovered as a special case.
 	
    	In the next section we will apply this fractional conservation of mass in one dimension for modelling
    	 the real gas flow through an highly heterogenous porous medium.
    	
    \section{Derivation of real gas diffusivity equation within the fractional conservation of mass}
    
    The main aim of this note is to reconsider classical models of real gas propagation through heterogeneous porous media in light of the fractional conservation of mass.  
    We will assume that the fluid flow is for simplicity in one dimension and in isothermal condition. Moreover the permeability of the medium is assumed to be constant.
    Let us consider the fractional conservation of mass according to the approach of Wheatcraft and Meerschaert \cite{mark}. From equation \eqref{fmas} in the one-dimensional case, we have that 
    \begin{equation}\label{0}
    -\frac{\partial}{\partial t}\rho \phi = \frac{\partial^\gamma}{\partial x^\gamma}(v\rho), \quad \gamma\in(0,1).
    \end{equation}
    The velocity flux is given by the classical Darcy law
    \begin{equation}\label{1}
    v=-\frac{k}{\mu}\frac{\partial p}{\partial x},
    \end{equation}
    where $\mu$ is the fluid viscosity coefficient and $k$ the permeability. 
    
    It is well-known that the Darcy law is a purely empirically-based law and it works well in many realistic models of fluid flow in porous media. Despite this, in many other physical cases, some modifications of the Darcy law have been successfully applied, such as the Forchheimer Law. Moreover space or time-fractional generalizations of the Darcy law have been recently studied, for example in \cite{cap} and \cite{physica d}. 
         The role of these more general forms of the Darcy law in the context here considered, should be object of further research.
         
    In order to obtain a single equation governing the pressure field evolution, we should consider the relation between density and pressure field that is given by the following equation of state for real gas 
    \begin{equation}\label{2}
    \rho = \frac{M}{RT}\frac{p}{z},
    \end{equation}
    where $M$ is the molecular weight of the gas, $R$ is the universal gas constant, $T$ the absolute temperature and $z$ the so-called \textit{gas deviation factor}. We recall that the gas deviation factor
    $z$ is by definition the ratio of the volume actually occupied by the gas at 
    a given pressure and temperature, to the volume occupied if it behaved ideally.
    In the more general case the gas deviation factor might depend by temperature and pressure.
    Typically the gas deviation factor is close to 1 (i.e. the gas behaves as an ideal gas) at low pressures and high temperatures, while for high pressure the gas is said to be \textit{super-compressible}.
    
    Substituting equations
    \eqref{1} and \eqref{2} in \eqref{0}, we obtain
    \begin{equation}\label{uno}
    \frac{\partial}{\partial t}\left(\phi \frac{p}{z}\right)=
    \frac{\partial^\gamma}{\partial x^\gamma}\left(\frac{k}{2\mu z}\frac{\partial p^2}{\partial x}\right).
    \end{equation}
    
    Since both the porosity and the gas deviation factor are pressure-dependent, we observe that
    \begin{align}
    \nonumber &\frac{\partial}{\partial t}\left(\phi \frac{p}{z}\right)
    = \bigg\{\left(\frac{\partial\phi}{\partial p}\right)\frac{p}{z}+\phi \frac{\partial}{\partial p}\left(\frac{p}{z}\right)\bigg\}
    \frac{\partial p}{\partial t}\\
    &= \frac{\phi p}{z}(c_f+c_g)\frac{\partial p}{\partial t},
    \end{align}
    where
    \begin{equation}
    c_g= \frac{z}{p}\frac{d}{dp}\left(\frac{p}{z}\right)
    \end{equation}
    is the real gas compressibility and 
    \begin{equation}
    c_f = \frac{1}{\phi}\frac{d\phi}{dp}
        \end{equation}
    is the porous medium compressibility. Clearly these coefficients are constant only under specific assumptions on the pressure dependence of the porosity and gas deviation factor, in the general case they are pressure-dependent.
      
    Therefore, according to the fractional conservation of mass equation \eqref{0}, we finally arrive to the diffusivity equation for real gas flow in porous medium
    \begin{equation}\label{diff}
    \frac{\phi}{z}(c_f+c_g)\frac{\partial p^2}{\partial t}=
        \frac{\partial^\gamma}{\partial x^\gamma}\left(\frac{k}{\mu z }\frac{\partial p^2}{\partial x}\right).
    \end{equation}
    This is a nonlinear fractional partial differential equation hardly solvable because to the general pressure dependence of the porosity $\phi$ and of the compressibility coefficients. Since in the general case it is not possible to find exact analytical solutions to \eqref{diff}, in the literature it is frequently assumed that the gas deviation factor $z$ and the diffusivity coefficient $\alpha = k/(\mu\phi(c_f+c_g))$ are approximately constant. This is a strong physical assumption, that is generally valid only for short times, considering the physical parameters appearing in the diffusivity coefficient evaluated for the mean pressure over a time interval. In practice this diffusivity coefficient is frequently considered in correspondence to the initial reservoir pressure $p_i$.  \\
    Under this assumption, the non-linear fractional equation \eqref{diff} is linearizable by means of the simple change of variable $u(x,t):= p^2(x,t)$. In this case we therefore arrive to the linear space-fractional diffusion equation involving the space-fractional Caputo derivative of order $\gamma+1$
    \begin{equation}\label{diffu}
    \phi(c_f+c_g)\frac{\partial u}{\partial t}=\frac{k}{\mu }
            \frac{\partial^{\gamma+1}u}{\partial x^{\gamma+1}}.
    \end{equation}
    This is one of the few cases of nonlinear fractional equations arising from physical models that are $C$-integrable, \textit{i.e.} integrable by change of variable (see \cite{io} and the references therein).
    Equation \eqref{diffu} is widely studied in the mathematical literature, we refer for example to \cite{fra} for a complete treatment.\\
   It is well-known in the physical literature that this equation arises in the context of anomalous diffusion processes in models of particles dynamics where long jump distributions are considered (see for example \cite{metz}). Anomalous diffusions include a wide class of processes whose variance does not grow linearly in time, in
   contrast to normal diffusion. \\
   Anomalous diffusion processes related to space-fractional diffusion equations, have found relevant applications in different fields of applied sciences, including particle advection-diffusion on Earth surface \cite{schumer}. \\
   As already seen, in equation \eqref{diff} we assume all the physical coefficients appearing in the diffusivity $\alpha$ as constant. 
   In the next section we evaluate the role played by a power-law pressure dependent permeability in the pressure field diffusivity equation.
   In this case we are able to find exact solutions, even if the obtained equations have the form of space-fractional porous medium-type equations. In Appendix C we provide a simple mathematical scheme to reduce a more general case to the space-fractional porous-medium type equation.

    \section{Diffusivity models with permeability variations induced by variations of pressure}
    
    In realistic models of gas flow through porous media, the permeability coefficient depends by variations of pressure, and
    for high pressure cases an empirical power-law 
    dependence is observed (see for example \cite{shapiro} and the references therein)
	\begin{equation}\label{kappa}
	k(p)\sim k_0 \ p^\beta, \quad \beta > 0.
	\end{equation}    
	Taking into account this dependence implies that higher order nonlinearities appear in the diffusivity equation governing
	the fluid flow in heterogeneous medium. Indeed, considering the governing equation \eqref{diffu} and the pressure-dependence 
	of the porosity \eqref{kappa} we obtain the equation
	\begin{equation}\label{diff1}
	    \phi(c_f+c_g)\frac{\partial p^2}{\partial t}=\frac{k_0}{\mu }
	        \frac{\partial^\gamma}{\partial x^\gamma}\left(p^\beta\frac{\partial p^2}{\partial x}\right).
	    \end{equation}
	In this case the substitution $p^2(x,t):= u(x,t)$ leads to a space-fractional porous-medium-type nonlinear equation
	\begin{equation}\label{bar}
	 \phi(c_f+c_g)\frac{\partial u}{\partial t}=\frac{k_0}{\mu }
		        \frac{\partial^\gamma}{\partial x^\gamma}\left(u^{\beta/2}\frac{\partial u}{\partial x}\right).
	\end{equation}
	Fractional porous medium type equations have been widely studied in the recent mathematical literature, even if with a
	quite different formulation and starting from different assumptions (see e.g. \cite{vaz} and references therein).
	Here we have considered a new approach based on the physical assumptions leading to the fractional conservation of mass as discussed in \cite{mark}.
	As an outcome of our analysis, we have found that a fractional porous-medium type equation arises also in the treatment of diffusivity equation of gas flow in higly heterogeneous media, where the permeability variations are induced by changes of pressure. As far as we know the space-fractional equation \eqref{diff1} was not studied before. The second part of this paper is devoted to the analysis of some classes of
	explicit solutions by using generalized separating variable and invariant subspace methods. It is well known that exact solutions of nonlinear
	evolution equations play an important role for the study of relevant features like the aysmptotic behavior, finite velocity of propagation or
	blow up in finite time. Few exact results for nonlinear evolution equations involving fractional derivatives in space or time are present in the
	literature. In this field of research the applications of Lie symmetry methods and invariant subspace method play a central
	role as it is proved by many recent publications such as \cite{noi,gazi,saha}. The full mathematical discussion about the properties of the equation \eqref{bar} is beyond the
	aims of this paper.  However we will show that in an interesting particular case Barenblatt--type solutions can be obtained.

	\subsection{Exact analytical results}
	
	\subsubsection{Stationary solutions}
	
	 As a first simple solution that can be analitycally investigated, we study the stationary solutions of equation \eqref{bar}. This means that
	 we should solve the following fractional nonlinear ordinary differential equation
	 \begin{equation}\label{bar1}
	\frac{k_0}{\mu }
	 		        \frac{d^\gamma}{d x^\gamma}\left(u^{\beta/2}\frac{d u}{d x}\right)= 0.
	 \end{equation}
	 A solution of equation \eqref{bar1} is given by 
	 \begin{equation}
	 u(x)= (c_1+c_2 x)^{\frac{2}{\beta+2}},
	 \end{equation}
	 where the real constants $c_1$ and $c_2$ depends by the boundary conditions.
	For the physical non-negativity constraint, we must consider suitable constants and the restriction $x>0$.
	The corresponding pressure profile will be given by 

	\begin{equation}\label{ssolu}
	p(x)= \sqrt{(c_1+c_2 x)^{2/{(\beta+2)}}}.
	\end{equation}
	For example, considering the steady propagation in the bounded domain 
	$x\in[0,L]$, with $L>0$ and taking the boundary conditions $p(0)= p_0>0$
	and $p(L)= p_L$, we have that the constants appearing in \eqref{ssolu}
	are given by
	\begin{align}
	\nonumber &c_1= p_0^{\beta+2},\\
	\nonumber &c_2= \frac{p_L^{\beta+2}-p_0^{\beta+2}}{L}.
	\end{align}
	and therefore the pressure field evolves according to the following equation
	\begin{equation}
	p(x)= \sqrt{\left(p_0^{\beta+2}+\frac{p_L^{\beta+2}-p_0^{\beta+2}}{L} \ x\right)^{2/{(\beta+2)}}}.
	\end{equation}
	
	\subsubsection{Translating front solution}
    It is simple to prove that equation \eqref{bar} admits a separating variable solution (see \cite{poly}) of the form 
	\begin{equation}\label{sepva}
	u(x,t)= \mathcal{X}(x)+\mathcal{T}(t).
	\end{equation}
	
	Indeed by substituting \eqref{sepva} in \eqref{bar}, we should solve the following ordinary differential equations
	\begin{equation}
	\begin{cases}
	&\displaystyle{\frac{d\mathcal{T}}{dt}= \lambda, \quad \lambda \in \mathbb{R},}\\
	& \alpha\displaystyle{\frac{d^\gamma}{dx^\gamma}\chi^{\beta/2}\frac{d\chi}{dx}= \lambda,} \quad \alpha= \frac{k_0}{	\mu \phi(c_f+c_g)}
	\end{cases}
	\end{equation}
	whose trivial solutions are given by 
	\begin{align}
	\nonumber &\mathcal{X}(x)=C_1 \ x^{\frac{2(\gamma+1)}{\beta+2}} \\
	\nonumber &\mathcal{T}(t)= \lambda t+C_2,
	\end{align}
	where 
	\begin{equation}
	C_1 = \bigg[ \frac{\beta+2}{2(\gamma+1)}\frac{\lambda}{\alpha\Gamma(\gamma+1)}\bigg]^{\frac{2}{\beta+2}}
	\end{equation}
	and the real constant $C_2$ depends by the initial condition. This solution corresponds to a rigid translation
	of the initial profile $u(x,0)\propto C_2+ x^{\frac{2(\gamma+1)}{\beta+2}}$.
	
	Therefore, going back to the equation governing the evolution of the pressure field, assuming as initial condition 
	\begin{equation}
	p(x,0)= \sqrt{p_0+C_1 \ x^{\frac{2(\gamma+1)}{\beta+2}}},
	\end{equation} 
	a solution to \eqref{bar} is given by
	\begin{equation}
	p(x,t)= \sqrt{p_0+C_1 \ x^{\frac{2(\gamma+1)}{\beta+2}}+\lambda t}.
	\end{equation} 
	This is a rigid translation without changing of shape.
	
	\subsubsection{Separating variable solutions}
	A second interesting class of analytical solutions of porous-medium-type equations is given by separating variable solutions, i.e. (see the encyclopedic Handbook of Polyanin and Zaitsev \cite{poly}) 
	\begin{equation}\label{see}
	u(x,t)= X(x)T(t),
	\end{equation}
	where the functions $X(x)$ and $T(t)$ should solve the following nonlinear ordinary equations
	\begin{align}\label{spec}
	&\phi(c_f+c_g)\frac{dT}{dt}= \lambda T^{\frac{\beta}{2}+1}, \\ &\frac{k_0}{\mu} \frac{d^\gamma}{dx^\gamma} X^{\beta/2}\frac{dX}{dx}=
	\lambda X.\label{speca}
	\end{align}
    This is a nonlinear eigenvalue probem and, as usual, this kind of solutions strongly depends by the value of $\lambda$. We consider the nonlinear eigenvalue problem \eqref{spec} for $\lambda>0$ that can be reduced to the more simple case $\lambda = 1$ by simple scaling of $X(x)$. The solution of the ordinary equation in $T(t)$ is rather trivial and it is given by
	\begin{equation}\label{ti}
	T(t)= \bigg[\frac{\beta}{2\phi(c_f+c_g)}(t_0-t)\bigg]^{-2/\beta},
	\end{equation}
	with $t_0>0$.
	The solution of \eqref{speca} clearly depends by the boundary conditions and in general is not trivial to find an explicit form.
	However it is simple to prove that 
	equation \eqref{speca} admits as a solution the following function
	\begin{equation}\label{ix}
	X(x) = \left(\frac{\mu}{k_0}\frac{\beta}{2\gamma+2}\frac{\Gamma(\frac{2\gamma+2}{\beta}+1)}{{\Gamma(\frac{2\gamma+2}{\beta}+\gamma+1)}}\right)^{2/\beta}
	x^{\frac{2\gamma+2}{\beta}}.
	\end{equation}

	Therefore, substituting \eqref{ix} and \eqref{ti} in \eqref{see}, we obtain a simple solution by separating variables of the form
		\begin{equation}
		u(x,t)= \left(\frac{\mu}{k_0}\frac{2\phi(c_f+c_g)}{2\gamma+2}\frac{\Gamma(\frac{2\gamma+2}{\beta}+1)}{{\Gamma(\frac{2\gamma+2}{\beta}+\gamma+1)}}\right)^{2/\beta}
			\left(\frac{x^{2\gamma+2}}{(t_0-t)^2}\right)^{1/\beta}.
		\end{equation}
	This solution clearly leads to a blow-up in finite time for $t= t_0$ and should be carefully considered. This kind of solutions leading to an explosive behavior in finite time are interesting for the mathematical theory behind the space-fractional porous-medium equation \eqref{bar}.
	
	\subsubsection{Barenblatt-type solutions}
	
	A relevant role in the theory of porous-medium equations is played by the so-called Barenblatt solution, corresponding to the \textit{fundamental solution} of the porous medium equation, leading to a pressure profile propagating with finite velocity. We refer to the classical book of V\'azquez \cite{vazq} for a complete analysis about this topic.	\\
	Here we start our analysis, by considering self-similar solutions of \eqref{bar} of the form 
	\begin{equation}
	\mathcal{U}(x,t)= t^{-k}f(xt^{-s})= t^{-k}f(\eta), \quad \eta= xt^{-s},
	\end{equation}
	where $k$ and $s$ are similarity exponents and $f(\cdot)$ the self-similar profile.
	Since, by simple calculations we have that
	\begin{align}
		\nonumber & \frac{\partial \mathcal{U}}{\partial t}= -t^{-k-1}\left(k\ \mathcal{U}(\eta)+s \eta 
		\frac{\partial \mathcal{U}}{\partial \eta}\right)\\
		\nonumber &\frac{\partial^\gamma}{\partial x^\gamma}\mathcal{U}^{\beta/2}\frac{\partial \mathcal{U}}{\partial x} =
		t^{-s(\gamma+1)-\frac{k \beta}{2}}\frac{\partial^\gamma }{\partial\eta^\gamma}\mathcal{U}^{\beta/2}\frac{\partial \mathcal{U}}{\partial \eta}.
		\end{align}
	We have therefore 
	\begin{equation}
	-t^{-k-1}\left(k \ \mathcal{U}(\eta)+s \eta 
		\frac{\partial \mathcal{U}}{\partial \eta}\right)=	t^{-s(\gamma+1)-\frac{k \beta}{2}}\frac{\partial^\gamma }{\partial\eta^\gamma}\mathcal{U}^{\beta/2}\frac{\partial \mathcal{U}}{\partial \eta}.
	\end{equation}
	In order to eliminate the time dependence we have the first relation between similarity coefficients
	\begin{equation}\label{sim}
	k\left(\frac{\beta}{2}-1\right)+(\gamma+1)s = 1.
	\end{equation}
	We therefore should solve the nonlinear fractional '\textit{eigenvalue}'
	problem
		\begin{equation}\label{nlbar}
		\left(k \ f(\eta)+s \eta 
				\frac{\partial f}{\partial \eta}\right)+\frac{\partial^\gamma }{\partial\eta^\gamma}f^{\beta/2}\frac{\partial f}{\partial \eta}=0.
		\end{equation}
	The free parameter $\beta$  is then fixed on the basis of the physical constraints of the conservation of mass
	\begin{equation}
	\int \mathcal{U}(x,t)dx= const. 
	\end{equation}
	which implies $\alpha = s$ and, by substitution in \eqref{sim}, we have the explicit form of the similarity exponents
	\begin{equation}
	k = s = \frac{1}{\gamma+\frac{\beta}{2}+1}.
	\end{equation}
	In the general case, unfortunately, we are not able to solve the nonlinear fractional differential equation \eqref{nlbar}.
	 
	  We now show that, for a specific case of the model equation \eqref{bar}, we are able to find in explicit form a Barenblatt-type profile. For the more general case, this kind of solutions cannot be found by using the method here employed. A general analysis of the problem of finite velocity of propagation for this formulation of the fractional porous medium equation should be studied, but this is beyond the aim of this paper. In our view this is an interesting example that can be useful both for the physical meaning and the future mathematical analysis of the nonlocal porous medium equation \eqref{bar}. In this case we adopt the invariant subspace method \cite{gala} as a useful mathematical tool to find a Barenblatt-type solution. We refer to the monograph \cite{gala} and to the Appendix B for details about this method.\\
	Let us consider equation \eqref{bar} for $\beta = 2$, corresponding to the space-fractional Boussinesq equation. The resulting equation can be written as
	\begin{equation}\label{baren}
	\frac{\partial u}{\partial t}= \alpha \frac{\partial^\gamma}{\partial x^\gamma}\left(u\frac{\partial u}{\partial x}\right),
	\end{equation}
	with $(x,t)\in \mathbb{R}^+\times \mathbb{R}^+$ and the diffusion coefficient $\alpha= k_0/(\mu \phi(c_f+c_g))$. This is a non-local generalization of the classical Boussinesq equation that plays a key-role in hydrology. We observe that in the recent paper \cite{Bal}, the authors have
   discussed a physical derivation of the space-fractional Boussinesq equation for modelling unconfined underground flow.
   Let us define 
   \begin{equation}\label{baren1}
   F\left[t,u(x,t), \frac{\partial^\gamma u}{\partial x^\gamma}, \frac{\partial u}{\partial x}\right]:=\frac{\partial^\gamma}{\partial x^\gamma}\left(u\frac{\partial u}{\partial x}\right), \quad x\geq 0.
   \end{equation}
	It is simple to prove that the nonlinear operator $F[\cdot]$ defined in \eqref{baren1} admits as invariant subspace $W^2= \langle 1, x^{\gamma+1}\rangle$. \\
	Indeed by simple calculations (see Appendix A and B for more details), we have that
	\begin{equation}
	F(c_1+c_2 x^{\gamma+1})= (\gamma+1)\frac{\partial^\gamma}{\partial x^\gamma}\left(c_1 c_2 x^\gamma+c_2^2 x^{2\gamma+1}\right)= 
	(\gamma+1)\bigg[c_1c_2\Gamma(\gamma+1)+\frac{\Gamma(2\gamma+2)}{\Gamma(\gamma+2)}c_2^2 x^{\gamma+1}\bigg]
	\end{equation}
	and therefore the subspace $W^2= \langle 1, x^{\gamma+1}\rangle$ is invariant under the operator $F[\cdot]$.
	
	This means that we can search a solution of \eqref{baren} in the form 
	\begin{equation}\label{ss}
	u(x,t)= a(t)x^{\gamma+1}+b(t),
	\end{equation}
	By substituting \eqref{ss} into \eqref{baren}, we obtain
	\begin{equation}\nonumber
	x^{\gamma+1}\frac{da}{dt}+\frac{db}{dt}= \alpha\left(\frac{(\gamma+1)\Gamma(2\gamma+2)}{\Gamma(\gamma+2)}a^2 x^{\gamma+1}+
	\Gamma(\gamma+2)ab\right)
	\end{equation}
	and therefore the functions $a(t)$ and $b(t)$ must solve the coupled system of nonlinear differential equations
	\begin{equation}
	\begin{cases}
	&\displaystyle \dot{a}= \frac{\alpha(\gamma+1)\Gamma(2\gamma+2)}{\Gamma(\gamma+2)}a^2,\\
	&\dot{b}=\alpha\Gamma(\gamma+2)ab.
	\end{cases}
	\end{equation}
	By simple calculations we finally find the following solution
	\begin{equation}\label{solrot}
	u(x,t)= \frac{1}{t^{c_2}}\left[1-\frac{c_1 x^{\gamma+1}}{t^{1-c_2}}\right]_+, \quad x\geq 0
	\end{equation}
	where $(\cdot)_+ = \max\{\cdot,0\}$ (for the positivity physical constraint $u\geq 0$)
	and 
	\begin{equation}
	\begin{cases}
	\displaystyle c_1= \frac{\Gamma(\gamma+2)}{\alpha(\gamma+1)\Gamma(2\gamma+2)},\\
	\displaystyle c_2= \frac{\Gamma^2(\gamma+2)}{(\gamma+1)\Gamma(2\gamma+2)}.
	\end{cases}
	\end{equation} 
	This solution corresponds in the classical theory, to the \textit{source solution} that is related to a Dirac's delta function
	as an initial condition. In this case the pressure propagates with finite velocity, starting from a strong pulse and at fixed time the solution is always with compact support.
	The rigorous way to obtain this solution requests more sophisticated limit arguments and the development of the mathematical analysis of 
	weak solutions for this kind of problems. This is beyond the aims of this paper that is devoted to find some special solutions that can have a clear motivation in our physical framework.
	In our case, following the seminal investigations about the fundamental solutions of the porous medium equation, we have roughly cutted off the part of the profile that is physically meaningless (corresponding to negative values of pressure). This corresponds to find a pressure front propagating at finite velocity. 
	
	 \begin{figure}
	                             \centering
	                             \includegraphics[scale=.45]{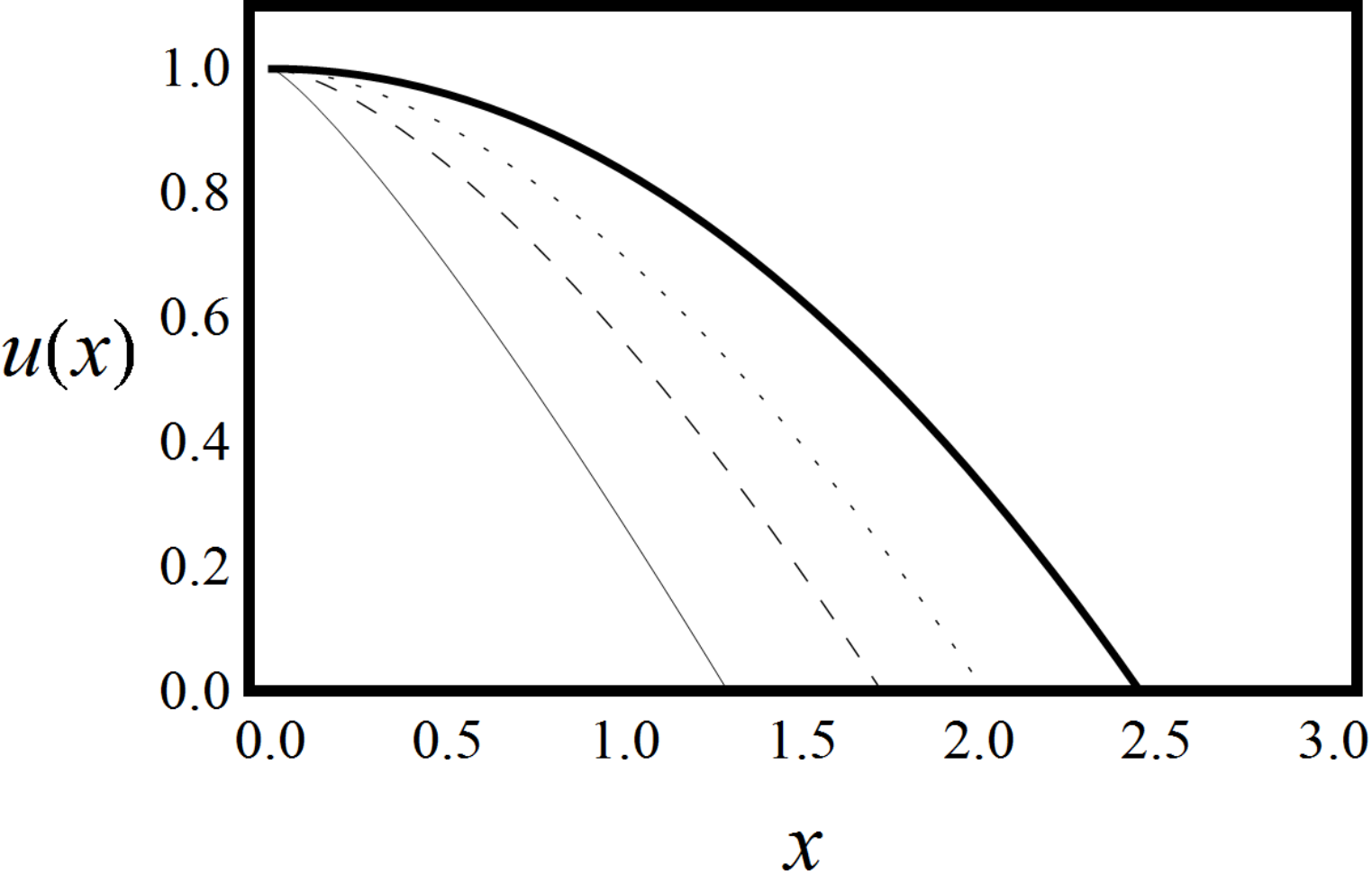}
	                             \caption{We represent the solution \eqref{solrot} for $x\geq 0$, and (for simplicity) $t=\alpha = 1$.
	                             The continuous
	           					line corresponds to $\gamma = 1/2$, the dashed to $\gamma = 1/5$, the dotted to $\gamma = 1/7$ and
	           					the bold line to $\gamma = 1$. The function is zero outside the set where the pressure is positive and therefore the support is compact.} 
	                             \label{figura0}
	                  \end{figure}
	This solution describes a front evolving as $x(t)\sim t^{\frac{1-c_2}{1+\gamma}}$, therefore with a strong dependence of the velocity propagation by the parameter $\gamma$, parametrizing nonlocal effects in the nonlinear evolution.\\
	We can also observe that for $\gamma = 1$, the classical Barenblatt solution for the porous medium equation is recovered (by choosing a suitable initial intensity of the source pulse). Indeed in this case we obtain that
	\begin{equation}
	u(x,t)= \frac{1}{t^{1/3}}\left[1-\frac{1}{6}\frac{|x|^2}{t^{2/3}}\right]_+,
	\end{equation}
	that coincides with equation (1.8) of \cite{vazq} up to a multiplicative constant. Observe that in our case the modulus is missing because we are considering the problem in the semiline $x\geq 0$. It is therefore clear the role played by the nonlocality that changes the shape and amplitude of the support of the evolving front, as can be seen in Fig.1. A relevant outcome of this result is related to the experimental evaluation of the fractional order of derivative and its physical meaning. Indeed, it can be found from the measured velocity of the front.\\
	This kind of solutions can play an interesting role also in the context of the studies about fluid-induced micoseismicity \cite{shapiro}, this will be object of further studies.
	We conclude this section, observing that the relevant feature of the finite velocity of propagation is therefore preserved in the space-fractional Boussinesq equation.

	\section{Conclusions}
	
	In this paper we have reconsidered the model of nonlinear diffusivity of real gas through a strongly heterogenous porous medium.
	In the derivation of the governing equations, we considered the recent discussions about the application of fractional derivatives
	in the classical conservation mass law to take into account power law flux or high heterogeneity of the medium. 
	We have therefore derived new classes of nonlinear nonlocal diffusivity equations. A part of this paper has been devoted to the derivation
	of rigorous exact analytic results for the obtained equations of space-fractional diffusion and porous-medium type.
	
	Nonlinear models involving space or time fractional derivatives should be still object of research in the physics of solid earth.
	This topic of research is motivated by recent discussions about the generalized Darcy law and the space-fractional conservation of mass in heterogeneous media.
	This paper is the first step in this direction, with the main aim to present the advantages of fractional calculus models for possible 
	future experimental validation and investigations. From a theoretical mathematical point of view, we suggest some issues that should be investigated, in particular regarding the existence of finite velocity propagating solutions and the more complicated multidimensional cases. 
		
	\section{Appendix A: Some details on fractional Caputo derivatives}
	
	Here we give some details about the fractional Caputo derivatives and some simple mathematical rules that we have applied in the text.
	Let $0 < \gamma\leq 1$, the Caputo fractional derivative is defined by
	\begin{equation}
	D_C^{\gamma}f(x)= 
	\begin{cases}
	\int_0^{x}\frac{(x-x')^{-\gamma}}{\Gamma(1-\gamma)}f^{'} (x') \, \mathrm dx', &, \gamma \in (0,1)\\
	f^{'}(x), & \gamma = 1,
	\end{cases}
	\end{equation}
	where $f^{'}(x)$ is the ordinary first order derivative with respect to $x$. We can observe that $D_C^{\gamma}f(x)= J_x^{1-\gamma}\left(\frac{df}{dx}\right)$, where 
	\begin{equation}
	J^{\gamma}_x f(x) = \frac{1}{\Gamma(\gamma)}\int_0^{x}(x-x')^{\gamma-1}f(x') dx',
	\end{equation}
	is the Riemann-Liouville integral of order $\gamma \in (0,1)$.\\
	It is simple to prove the following properties of Caputo fractional derivative of order $\gamma \in (0,1)$ (see e.g.\ \cite{kil}):
	\begin{align}
	\nonumber &D_C^{\gamma} J^{\gamma}_x f(x)= f(x),\\
	\nonumber &D^{\gamma}_C x^{\delta}= \frac{\Gamma(\delta+1)}{\Gamma(\delta-\gamma+1)}x^{\delta-\gamma} \qquad \delta \in (-1,0)\bigcup (0,\infty),\\
\nonumber &D^{\gamma}_C const. = 0.
	\end{align}

	\section{Appendix B: The invariant subspace method}
	
	The Invariant Subspace
	     Method, introduced in the literature by Galaktionov (see the monograph \cite{gala} for details),
	  allows to solve exactly nonlinear equations by separating variables.\\
	     We recall the main idea of this method: consider a scalar evolution equation
	     \begin{equation}\label{pros}
	     \frac{\partial u}{\partial t}= F\left[u, \frac{\partial u}{\partial
	     x}, \dots\right],
	     \end{equation}
	     where $u=u(x,t)$ and $F[\cdot]$ is a nonlinear differential
	     operator. Given $n$ linearly independent functions
	     $$f_1(x), f_2(x),....,f_n(x),$$
	     we call $W^n$ the $n$-dimensional linear space
	     $$W^n=\langle f_1(x), ...., f_n(x)\rangle.$$
	     This space is called invariant under the given operator $F[u]$, if
	     $F[y]\in W_n$ for any $y\in W_n$. This means that there exist $n$
	     functions $\Phi_1, \Phi_2,..., \Phi_n$ such that
	     \begin{align}
	     \nonumber &F[C_1f_1(x)+......C_n f_n(x)]= \Phi_1(C_1,....,C_n)f_1(x)+......\\
	     \nonumber&+\Phi_n(C_1,....,C_n)f_n(x),
	     \end{align}
	     where $C_1, C_2, ....., C_n$ are arbitrary constants. \\
	     Once the set of functions $f_i(x)$ that form the invariant subspace
	     has been determined, we can search an exact solution of \eqref{pros}
	     in the invariant subspace in the form
	     \begin{equation}
	     u(x,t)=\sum_{i=1}^n g_i(t)f_i(x).
	     \end{equation}
	     where $f_i(x)\in W_n$. In this way, we arrive to a system of ODEs.
	     In many cases, this problem is simpler than the original one and
	     allows to find exact
	     solutions by just separating variables \cite{gala}.
	     We refer to the monograph \cite{gala} for further details and applications of this method.
	     The first applications of the invariant subspace method to fractional equations is due to Gazizov and Kasatkin \cite{gazi}.
		
			\section{Appendix C: A mathematical analysis of a more general case}
			A strong limitation in the analysis developed in this paper is given by the assumption that the diffusivity coefficient (involving compressibility coefficients, porosity, permeability and viscosity) is constant and pressure-independent. Here we show that, under some mathematical assumptions, we are able to reduce a more general case to a fractional porous medium-type equation, similar to the one considered in the previous section. Here we provide a syntetic scheme to obtain this result:
			\begin{itemize}
			\item Assume that the coefficients appearing in \eqref{diff}
			are pressure-dependent according to the following equalities
			\begin{equation}
			\frac{\phi(c_f+c_g)}{z}\sim C_1 p^{n_1} \quad \frac{k}{\mu z}\sim C_2 p^{n_2},
			\end{equation}
			with $n_1,n_2, C_1,C_2$ arbitrary real positive constants. Therefore equation \eqref{diff} becomes			
			\begin{equation}\label{diffa}
			    C_1p^{n_1}\frac{\partial p^2}{\partial t}= C_2 \frac{\partial^\gamma}{\partial x^\gamma}p^{n_2}\frac{\partial p^2}{\partial x}.
			    \end{equation}
			\item Define $u = p^2$, such that \eqref{diffa} becomes
			\begin{equation}\label{diffa1}
			\frac{2C_1}{n_1+2}\frac{\partial}{\partial t}u^{\frac{n_1}{2}+1}= C_2\frac{\partial^\gamma}{\partial x^\gamma}u^{\frac{n_2}{2}}\frac{\partial u}{\partial x}.
			\end{equation}
			\item Define $s = u^{\frac{n_1}{2}+1}$ and therefore we finally obtain a space-fractional porous-medium-type equation
			\begin{equation}\label{pme}
			C_1\frac{\partial s}{\partial t}= 
			C_2\frac{\partial^\gamma}{\partial x^\gamma}s^{\frac{n_2+2}{n_1+2}-1}\frac{\partial s}{\partial x}
			\end{equation}
			Thus we can apply the methods used before to solve \eqref{bar} also in this more general case.
			
			\end{itemize}

    \end{document}